\documentclass{aa}

\usepackage{graphicx}
\usepackage{txfonts}

\newcommand{\as}{$\alpha$-stable }
\newcommand{\asb}{$\alpha$-stable}
\newcommand{\pdf}{\emph{pdf} }
\newcommand{\pdfs}{\emph{pdf}s }

\begin{document}

   \title{An \as approach to the
     study of the $P(D)$ distribution 
     of unresolved point sources in CMB sky maps}

%Using \as distributions to model the $P(D)$
%distribution of point sources in CMB sky maps}

%   \subtitle{}

   \author{D. Herranz
     \inst{1} \and
     E.E. Kuruo\u glu
     \inst{1}
     \and
     L. Toffolatti\inst{2}
   }
   
   \institute{Istituto di Scienza e Tecnologie dell'Informazione. \\
     CNR, Area della Ricerca di Pisa,
     via G. Moruzzi 1, 56124 Pisa, Italy. \\
     \email{diego.herranz@isti.cnr.it, kuruoglu@isti.cnr.it}
     \and
     Departamento de F\'\i sica, Universidad de Oviedo, \\ c/ Calvo
     Sotelo s/n, 33007 Oviedo, Spain
   }
   
   \date{Received 12 December 2003 / Accepted 8 April 2004}

   \abstract{
We present a new approach to the 
statistical study and modelling of
number counts of faint point 
sources in astronomical images, i.e. counts of sources
whose flux falls below the detection limit of a survey.
The approach is based on the theory of
\as distributions. We show that the non-Gaussian 
distribution of the intensity fluctuations produced by a generic
point source population -- whose number counts follow
a simple power law -- belongs to the \as family of distributions.
Even if source counts do not follow a simple power
law, we show that the \as model is still useful in many astrophysical
scenarios. With the \as model it is possible to totally describe
the non-Gaussian distribution with a few parameters
which are closely related to the parameters describing the source counts, 
instead of an infinite number of moments. Using
statistical tools available in the signal processing literature,
we show how to estimate these parameters in an easy and fast way.
We demonstrate that the model proves valid when applied to realistic
point source number counts at microwave frequencies. In the case of
point extragalactic sources observed at CMB frecuencies, our technique 
is able to successfully fit the $P(D)$ distribution of deflections 
and to precisely determine the main parameters which describe the
number counts. In the case of the Planck mission, the relative errors 
on these parameters are small either at low and at high
frequencies. 
We provide a way to deal with the presence of Gaussian noise
in the data using the empirical characteristic function of the
$P(D)$. 
The formalism and methods here presented can be very
useful also for experiments in other frequency ranges, e.g. X-ray
or radio Astronomy.
   }

\titlerunning{$\alpha$-stable modelling of the $P(D)$ distribution}

   \maketitle
%
%________________________________________________________________

\section{Introduction} \label{sec:intro}

The study of intensity (or temperature)
fluctuations in the Cosmic Microwave 
Background (CMB) radiation has become
one of the milestones of modern cosmology due to two main 
reasons. On the one hand, the 
angular power spectrum of the fluctuations allow us to place tight constraints
on the fundamental cosmological parameters
(see, e.g., Bennett et al.~\cite{bennetta}). 
For a recent review on the  study of CMB anisotropies, see 
Hu \& Dodelson (\cite{hu}).
On the other hand, the study of the
different physical sources (foregrounds) 
that contribute to the incoming radiation at microwave wavelengths
has a great scientific relevance in itself (De Zotti et al.~\cite{zotti2}). 
Therefore, a great deal of effort has been devoted to the task of
separating the different components that are present in CMB maps.
In general, component separation techniques take advantage of 
the statistical behaviours of each component to 
distinguish among them. Hence, it is important to have good
statistical models for each of the components under study. 

Among the different foregrounds that appear in CMB observations, 
extragalactic
point sources (EPS), i.e. indi\-vi\-dual galaxies whose typical 
angular size is
much smaller than the observing beam width (hence
the name `point sources') are especially difficult to deal with.
The galaxies that contribute to the observed total signal are
very diverse, corresponding to objects at different redshifts and with
different physical properties. This makes it impossible to establish a single
spectral behaviour for all of them, thus hampering the performance of 
methods that use multi-frequency observations 
to achieve an efficient component separation.
The spatial distribution of faint EPS is roughly uniform across the sky
even in presence of source clustering.
Therefore, Galactic cuts useful to avoid Galactic 
foregrounds such as synchrotron,
dust and free-free emission, are not applicable to avoid EPS contamination.

EPS contamination occurs when a set of point-like sources with
intensities distributed following a general law, usually modelled
as a power law, is observed by a detector using a given instrumental
response.
The final signal is a mixing in which the 
brightest sources are still individually detectable over a `confusion
noise' generated by the contributions of faint, unresolved sources;
this situation is very common in astronomical images
and it has been studied first at radio and X-ray frequencies.
The intensity (or temperature) 
distribution given by unresolved sources is strongly
non-Gaussian and shows long positive tails. 
This kind of behaviour is known in the signal 
processing literature as `impulsive noise'.

The effect of the confusion noise is two fold:
on one hand the mean value of this noise is positive, producing a `source
monopole' (integrated extragalactic background)
that has to be summed up to the other components and, on the other hand,
it gives rise to small scale intensity fluctuations.
At microwave frequencies, the fluctuations generated by undetected
sources can severely hamper the detection of true CMB anisotropies 
(Franceschini et al.~\cite{franceschini1}). Recently, 
Toffolatti et al. (\cite{T98};
hereafter T98) presented a detailed analysis of
the effect of point sources on CMB anisotropy maps.
By exploiting a cosmological evolution model for radio and
far-IR selected sources, they made precise predictions on source counts,
on confusion noise and on the angular power spectrum due to undetected
sources. In particular, they showed that the contribution of EPS
will be very relevant at the lower and higher frequency
channels of the future ESA Planck mission (Mandolesi et al.~\cite{mandolesi};
Puget et al.~\cite{puget}).
As for radio source counts, the predictions
of T98 have been confirmed by the first year data of the
NASA \emph{Wilkinson Microwave Anisotropy Probe} (WMAP) mission (Bennett
et al.~\cite{bennettb}), al least up to frequencies of $\sim 30 - 40$ GHz.
The WMAP all sky
catalogue of bright extragalactic sources (Bennett et al.~\cite{bennettb})
consists of 208 objects with fluxes $S>0.9\div 1$ Jy on a sky area of 10.38
sr ($\vert b^{II}\vert>10^\circ$) whereas the T98 model predicts 270-280
sources at 30 GHz in the same area with an average offset of $\sim 0.75$
between observed and predicted number of
extragalactic sources. Moreover, the distribution of energy
spectral indexes (i.e. $a$, $S\propto \nu^{-a}$) of sources in the WMAP
sample peaks around $a=0.0$, which is exactly the
mean value of the energy 
spectral index  adopted in T98 for ``flat''--spectrum sources,
i.e. the dominant source population at these frequencies
(the fraction of ``steep''--spectrum sources being
$\sim 10 \div 15 \%$).
It is also noticeable that the brightest source
detected by WMAP has a flux density of $S\simeq 25$ Jy which,
again, corresponds to the flux value for which the T98
model predicts 1 source all over the sky.
Another important result confirming the estimates of the T98 
cosmological evolution model is that a good agreement is currently found 
between predictions based on that model and data on the excess 
angular power spectrum at small angular scales as well as on the angular 
bispectrum detected in the WMAP Q and V bands 
(Komatsu et al.~\cite{komatsu}, Arg\"ueso et al.~\cite{paco}).
Two other independent samples of extragalactic sources at 31 and 34
GHz -- from CBI (Mason et al.~\cite{mason}) and VSA (Taylor 
et al.~\cite{taylor})
experiments, respectively -- show that the T98 model correctly
predicts number counts down to, at least, $S\simeq 10$ mJy.
Therefore, we can confidently use the T98 model for simulating
Poisson distributed EPS in CMB sky maps, at least up to 50-100
GHz. At higher frequencies, more recent models can give a better fit to
current data on source counts (see section~\ref{sec:dusty}).
As the outcomes of the methods to be presented here
are \emph{model independent}, we still use the T98 model
throughout the paper.

The current low sensitivity of detectors at CMB frequencies makes it
impossible to test directly model counts down to fainter fluxes.
On the other hand, more information on counts of faint sources,
i.e. sources with fluxes fainter
than the detection threshold of a given experiment, can be extracted by the
analysis of the intensity fluctuations of point sources.
The probability density function \pdf of 
fluctuations due to undetected point sources,
as a first step to the modelling of the confusion noise,
has been studied since the middle of the last century 
(Scheuer~\cite{scheuer1},~\cite{scheuer2};
Condon~\cite{condon}; Hewish~\cite{hewish}). 
These works have shown that it is possible to find analytical
expressions for the characteristic function of the 
\pdf
(in Fourier
space), but not for the \pdf in the real space.
This fact has hampered the development of specific statistical tools to deal
with the EPS confusion noise. 

In analysing a sky map there are two traditional ways of determining
the main statistical properties of a given source population. One possibility
is to detect the brightest point sources in a given data set, e.g. using a
linear filter to detect them, and then obtain parameters such as the number 
counts, their slope, etc. For example, Vielva et al. (\cite{vielva}) 
detect point sources
in realistic Planck simulations using a Mexican Hat Wavelet technique and
compare the number of detections with the input number counts, which
correspond to the T98 model. The other possibility is to directly study
the \pdf of the confusion noise which, in general, is mixed
with the signal coming from CMB and the other foregrounds plus instrumental
noise. This is generally performed using statistical indicators such
as the moments up to a certain degree (see, e.g., Rubi\~no-Mart\'\i n \&
Sunyaev~\cite{rubi} and Pierpaoli~\cite{pierpaoli}) or the 
non-Gaussianity of the wings.
A computationally more complex way is to calculate numerically
the theoretical \pdf assuming some model for source counts and trying
to fit it to the data (Condon \& Dressel~\cite{condon2};
Franceschini et al.~\cite{franceschini1}).
Anyway, the lack of an analytical form for the \pdf
makes difficult to establish the optimal
estimator of its parameters. In particular, it is
not clear how many moments are necessary to 
characterise the \pdf (in principle, infinite of them) or which ones
are more appropriate for extracting
information (it is generally assumed,
on the basis of mere intuition, that the third order
moment should be one of them).

In this paper we will focus on the application of a novel 
formalism, the \as distributions, to model the 
 \pdf of the intensity fluctuations due to point
extragalactic sources.
\as distributions are known to be very efficient in modelling 
impulsive noise. 
They have a number of interesting mathematical properties that make them
very attractive; in particular, it is possible to show that the Gaussian
distribution is a 
special case of the more general class of \as distributions
and that \as distributions
satisfy a generalised form of the central limit theorem. Moreover,
in this work we show that the \pdf of a theoretical
power law representing the number counts of extragalactic sources
observed with a filled-aperture instrument must
follow exactly an \as distribution.\footnote{The same analysis can also be
applied to images obtained by interferometry techniques (insensitive to the
zero level). In this case, the resulting \pdf shows 
the same shape as for filled-aperture images but it is centred at 0.
If filled-aperture measurements are currently performed by means of
dual-beam scans, subtracting one signal form the other, filled-aperture and
interferometry techniques yield similar \pdfs (see, e.g.,
De Zotti et al.~\cite{zotti1}).}
The great advantage is that \as distributions are completely described
by a small number of parameters, what makes them relatively easy to deal with.
Optimal techniques 
already existent in the signal processing literature are easy to adapt
to directly extract the main parameters of the source number counts
(namely, the slope of the number counts power law and its normalisation)
without having to resort to clumsy statistics. 
Finally, the methods can be generalised for dealing with mixtures
of signals, as is the case when the EPS population is added to
Gaussian instrumental noise.

Therefore, this work aims at four objectives: first, we will
demonstrate that a generic extragalactic point source population 
whose number counts
follow a power law inevitably leads, when observed with a filled-aperture
instrument, to a \pdf which
belongs
to the family of \as distributions.
Then, we will devote some time to introduce the \as distributions
to the community,
reviewing their main properties and 
giving the necessary references for further reading. 
In a third part of this paper
we will discuss how \as distributions can be used to 
model and study more realistic cases, such as the truncated power
law, and we will show that the model is a good approximation for most
cases of astronomical interest. Finally, we will discuss the case
in which other astronomical or instrumental signals --such as
the Cosmic Microwave Background (CMB) radiation or instrumental
noise-- are mixed with the point sources. In that case, a method will
be suggested that is able to obtain the parameters of the point
source deflection distribution even in presence of `contamination'.

The structure of this paper is as follows: in Sect.~\ref{sec:pdf}
we review the basics of the derivation of the characteristic 
function of the deflection distribution.
In Sect.~\ref{sec:alpha} the \as distributions and their 
main properties are introduced.
 Section~\ref{sec:parameter} deals with the extraction 
of the physical parameters of the
EPS population using the \as formalism. In  Sect.~\ref{sec:Planck} 
we study the application of
the formalism to more realistic 
source models, using as an example the T98 point source model.
Parameter estimation of \as processes mixed with Gaussian
noise is considered in section~\ref{sec:mixing}.
A few considerations about the implementation of these techniques
for the future Planck mission are given in section~\ref{sec:future}.
Finally, in  Sect.~\ref{sec:conclusions}
we summarise our conclusions.

%__________________________________________________________________

\section{Source counts and the deflection 
probability function $P(D)$} \label{sec:pdf}

Let us consider a population of EPS whose differential
number counts can be described in a power law form:
\begin{equation} \label{eq:powlaw}
n(S)=kS^{-\eta}, \ \ \ S>0,
\end{equation}
\noindent
where $\eta$ is the \emph{slope} of the differential counts power law, $k$
is called its \emph{normalisation} and $S$ is the intrinsic flux.
The sources are assumed to be distributed uniformly across the 
sky and, at the moment,
we will assume that eq. (\ref{eq:powlaw}) holds for all $S>0$. The
 sources are now observed
with an instrument whose angular response is $f(\theta,\phi)$, 
not necessarily normalised to
unity at the peak. Then, the mean number of sources responses
of intensity $x=f(\theta,\phi)S$ in the beam at any time is
\begin{equation} \label{eq:r}
R(x) = 
\int n\left[\frac{x}{f(\theta,\phi)}\right]\frac{d\Omega}{f(\theta,\phi)}.
\end{equation}
Substituting eq. (\ref{eq:powlaw}) into eq. (\ref{eq:r}) we have that
\begin{equation} \label{eq:analr}
R(x)=k\Omega_e x^{-\eta},
\end{equation}
where
\begin{equation}
\Omega_e=\int \left[ f(\theta,\phi) \right]^{\eta-1} d\Omega
\end{equation}
\noindent
is a geometrical factor called \emph{effective beam solid angle}. 
Let us now define the 
\emph{deflection} $D$ as the fluctuation field that is observed, 
that is $D=I-\langle I \rangle$,
where $I$ is the intensity at a given point (time) and $\langle I 
\rangle$ is its average value, i.e. $\langle I\rangle$ represents the
extragalactic background due to undetected EPS.
Let us define the characteristic function of a given 
function $g(x)$ as
\begin{equation} \label{eq:charf2}
G(w) = \int_{-\infty}^{\infty} g(x) \ e^{-iwx} dx.
\end{equation}
Scheuer (\cite{scheuer1}) showed that the characteristic function 
of the  probability distribution $P(D)$ is
related to the characteristic function of $R(x)$ through
\begin{equation} \label{eq:charf1}
\psi(w)=\exp \left[ r(w)-r(0) \right],
\end{equation}
where $\psi(w)$ and $r(w)$ are the characteristic 
functions of $P(D)$ and $R(x)$, respectively. 

Using eqs. (\ref{eq:analr}), (\ref{eq:charf2}) and 
(\ref{eq:charf1}) it is possible to
calculate the characteristic function $\psi(w)$.
This calculation has been performed by several authors,
including Scheuer (\cite{scheuer1}), Condon (\cite{condon}),
Barcons (\cite{barcons}) and Franceschini et al. (\cite{franceschini1}). 
After some effort we obtain
\begin{equation} \label{eq:charf3}
\psi(w)=\exp \left\{ i \mu w - \gamma \left| w \right|^{\alpha}
\left[ 1  + i \beta \textrm{sgn}(w) \tan
 \left( \frac{\alpha \pi}{2} \right) \right] \right\},
\end{equation}
\noindent
where the parameters $\alpha$, $\beta$, $\gamma$ 
and $\mu$ relate to the physical 
parameters of the EPS and of the detector through
\begin{equation} \label{eq:alpha}
\alpha  =  \eta -1 ,
\end{equation}
\begin{equation} \label{eq:beta}
\beta   =  \frac{1}{\pi} \Gamma \left(\frac{1+\alpha}{2}\right)
\Gamma \left(\frac{1-\alpha}{2} \right)
\cos \left(\frac{\alpha \pi}{2}\right) = 1 ,
\end{equation}
\begin{equation} \label{eq:gamma}
\gamma  =  \frac{\pi^{3/2} k 
\Omega_e}{2^{\alpha+1}\Gamma\left(\frac{\alpha+1}{2}\right)
\Gamma\left(\frac{\alpha+2}{2}\right)\sin\left(\frac{\alpha \pi}{2}\right) } ,
\end{equation}
\begin{equation} \label{eq:mu}
\mu   =  \frac{k \Omega_e}{1-\alpha} \lim_{a \rightarrow 0^+} a^{1-\alpha}.
\end{equation}

The terms in eq. (\ref{eq:charf3})
have been arranged in this way for a reason that will be 
clear in the following.
The second equality in eq. (\ref{eq:beta}) is due to the 
properties of the gamma function (Abramowitz~\cite{abramowitz}),
and as far has we know it hasn't been noticed before now. 
The previous equations are valid for $1 < \eta < 3$. For $\eta > 2$ 
the parameter $\mu$ is not finite, a situation
equivalent to the classic Olbers' paradox
in which the observed integrated flux density 
is infinite in all directions of the sky.

Equation (\ref{eq:charf3})
has an important drawback: to obtain the \pdf of the deflections, $P(D)$, 
it is necessary to 
make the inverse Fourier transform
of $\psi(w)$ which, in general, 
cannot be evaluated analytically. Although
it can be performed numerically,
the computational cost can be high if many different 
realisations are needed for a particular task,
and numerical integration does not lead to closed form
solutions. Instead of doing that, 
let us see what can
be learnt  from the characteristic function itself.
As we will see in the next section, eq. (\ref{eq:charf3})
corresponds exactly to the actual definition of
a family of distributions that is known in the statistical
signal processing literature as \emph{$\alpha$-stable distributions}.

%__________________________________________________________________

\section{A short introduction 
to $\alpha$-stable distributions} \label{sec:alpha}

In this section we will make a summary introduction to 
the \as distributions, in order to clarify some concepts that will
be used later. Starting from the particular case of the EPS $P(D)$
characteristic function in
eq. (\ref{eq:charf3}), that we will
see that belongs to the \as family, we will proceed to a more
general description of \as phenomena that could be useful 
for astronomers.

\begin{figure} 
\centering
\includegraphics[width=\columnwidth]{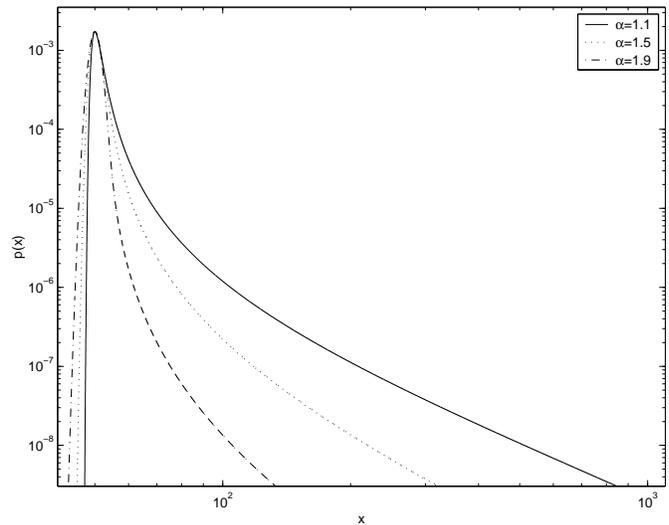}
\caption{Probability density functions 
corresponding to three different $\alpha$-stable 
models.
 The parameters of the \as
are $\beta=1$, $\gamma=1$ and
$\alpha = $ 1.1 (solid line), 1.5 (dotted line) and
1.9 (dot-dashed line). 
For an easier comparison among them, the $\mu$ parameter 
of each distribution
has been set so that
the maximum of all the \pdfs coincide at the arbitrary value $x=50$. 
The \pdf becomes more impulsive
(departs more from the Gaussian case) as $\alpha$ decreases.}
\label{fig:fig1}
\end{figure}

The characteristic function in eq. (\ref{eq:charf3}) corresponds  
to a \pdf that in general
should be calculated numerically and that 
exhibits heavier tails than a Gaussian
distribution. 
In Fig.~\ref{fig:fig1} the \pdf
corresponding to eq. (\ref{eq:charf3}) with $\beta=1$, $\gamma=1$ and three 
different values of
$\alpha$ (1.1, 1.5 and 1.9) are shown.
 The presence of heavy tails mean that `glitches' are more
likely to occur than in the Gaussian case. In the signal processing 
literature, 
probability 
density functions with tails heavier than the Gaussian
are called to be \emph{impulsive}.
A process is impulsive if it takes large values
that significantly deviates from the mean value
with non-negligible probability. 
These large values often appear as conspicuous outlayers. 
Impulsive processes are ubiquitous in many
`real-world' problems,
from atmospheric noise caused by electric discharges to financial
 time series data.
For more information on impulsive noise, see Kuruo\u glu (\cite{ercan1}).
A great deal of effort has been done to model impulsive processes; 
is in that
context that the \as distributions have
experienced popularity.
Other models that deal with impulsive processes, such
as the Middleton's (Middleton~\cite{middleton}), 
Cauchy and Student-T models, are very
{\it case specific} while the \as 
model is general and has a strong theoretical justification.

Although \as distributions were known from the beginnings of XX$^{th}$
century
(L\'evy~\cite{levy}), it was not until the 
work of Shao \& Nikias (\cite{shao}) that 
they received more interest in the signal processing literature. 
The \as distribution
is a generalisation of the Gaussian distribution 
that furnishes tractable examples of impulsive behaviour
and allows us to describe such behaviour by means of a small number
of parameters. The \as distributions are usually defined by their
characteristic function:
\begin{equation} \label{eq:alphastable0}
\psi(w)= \exp \left\{ i \mu w - \gamma \left| w 
\right|^{\alpha} B_{w,\alpha} \right\},
\end{equation}
\begin{equation} \label{eq:alphastable}
B_{w,\alpha} =
\left\{
\begin{array}{ll}
\left[ 1  + i \beta \textrm{sgn}(w) \tan \left( \frac{\alpha \pi}{2}
 \right) \right] &
\textrm{if $\alpha \neq 1$} \\
\left[ 1  + i \beta \textrm{sgn}(w) \frac{2}{\pi} \log{\left|w\right|}
 \right] &
\textrm{if $\alpha = 1$} \\
\end{array}
\right .
\end{equation}
\noindent            
where $-\infty < \mu < \infty, \ \gamma > 0, \ 0 < \alpha \leq 2$ 
and $-1 \leq \beta \leq 1$.
The four parameters $\mu, \ \alpha, \ \beta$ and $\gamma$ uniquely and
 completely 
determine the stable distribution.
The meanings of these parameters are:
\begin{enumerate}
\item The parameter $\alpha$ is called the \emph{characteristic exponent} 
and sets the
degree of impulsiveness of the distribution. For $\alpha=2$ the 
distribution corresponds 
to the Gaussian distribution and, as $\alpha$ decreases the distribution
gets more and more impulsive.
Another particular case is when $\alpha=1$ and $\beta=0$,
 that corresponds to the Cauchy distribution.
 For $\alpha \notin (0,2]$ the inverse Fourier transform of
$\psi(w)$ is not positive-definite and hence is not a proper probability
density function.
%%%%%%%%%%%%%%%%%%%%%%%%%%%%%%%%%%%%%%%%%%%%%%%%%%%%%%%%%%%%%%%%%%%%%%%%
%% Esta frase no se si ponerla: es cierto lo que digo, pero hace mas endeble
%% el metodo...podemos evitar ponerla...solo se anhade si la pide el referee,
%% de forma explicita!
%%
%% "This means that \as distributions cannot be applied when
%% $\eta> 3.$. Undoubtedly, this is a gap of the method given that
%% the slope of the differential counts can be $>3$ in some
%% frequency interval."
%%%%%%%%%%%%%%%%%%%%%%%%%%%%%%%%%%%%%%%%%%%%%%%%%%%%%%%%%%%%%%%%%%%%%%%%%%%%
\item The parameter $\beta$ is called \emph{symmetry 
parameter} and determines the skewness of
the distribution. Totally symmetric distributions have 
$\beta=0$, whereas $\beta = \pm 1$
corresponds to totally skewed distributions.
\item The parameter $\gamma$ is called \emph{scale parameter}. 
It is a measure of the spread
of the samples from a distribution around the mean. 
When $\alpha=2$ we get the Gaussian 
case and then $\gamma=\sigma_G^2/2$, where $\sigma_G$ 
is the dispersion of the Gaussian.
\item The parameter $\mu$ is called \emph{location parameter} 
and basically corresponds
to a shift in the x-axis of the \emph{pdf}.  For a symmetric ($\beta=0$) 
distribution, 
$\mu$ is the mean when $1 < \alpha \leq 2$ and the median 
when $0 < \alpha \leq 1$.
\end{enumerate}
As it can be seen, the first case in eqs. (\ref{eq:alphastable0}) and 
(\ref{eq:alphastable}), $\alpha \neq 1$, has exactly
the same expression as eq. (\ref{eq:charf3}). 
For simplicity, 
through this paper we are not
going to consider the case $\alpha=1$, as it corresponds
to a single point (of zero measure) in the interval $(0,2]$.

Going back to the expressions in Sect.~\ref{sec:pdf}, we see that
eq. (\ref{eq:charf3}) held when $1 < \eta < 3$, that is, when $0 < \alpha < 2$.
As $k$ and $\Omega_e$ are positive in eq. (\ref{eq:gamma}) and $0 < \alpha < 2$
we have that $\gamma > 0$. By eq. (\ref{eq:beta}) we know that $\beta=1$.
Therefore, eq. (\ref{eq:charf3}) is exactly the 
\emph{characteristic function} of
 an \as distribution with maximum positive skew.

The fact that a population of point sources that are distributed in
intensity following a power law and that are observed with a pencil-beam
instrument produce an \as distribution of deflections is very convenient. 
The \as representation offers several advantages:
\begin{enumerate}
\item \emph{Simplicity}: a non-Gaussian distribution that follows
eq. (\ref{eq:alphastable}) can be completely described by only 
four parameters,
instead of an infinite number of moments.
\item \emph{Mathematical justification}: \as distributions include 
as a particular
case the Gaussian distribution, and share with it many desirable 
properties. First,
they satisfy the \emph{generalised central limit theorem} 
which states that the limit
distribution on infinitely many i.i.d. random variables, 
possibly with infinite
variance distribution, is a stable distribution (Feller~\cite{feller}). 
Therefore, the
use of \as distributions is strongly justified from the 
theoretical point of view,
as they are able to describe a wider range of data 
which might not satisfy the classical
central limit theorem. In second place, \as distributions 
have the \emph{stability}
property: the output of a linear system in 
response to \as inputs is again \as and
various aspects of linear system theory developed for Gaussian 
signals extend directly
to the case of signals with \as distribution.
For more information on the mathematical foundation of stable distributions,
see  
Samorodnitsky \& Taqqu (\cite{samorodnitsky})
 and references therein.
\item \emph{Ubiquity}: \as can be shown to be the limit distribution of natural
noise processes under realistic 
assumptions pertaining to their generation mechanism and
propagation conditions (Nikias \& Shao~\cite{nikias}). 
They agree with empirical data extremely well 
in so different situations as noise in telephone 
lines, atmospheric noise, radio networks,
radar systems, financial time series, etc. 
Even in cases in which there is not a strong theoretical or physical
evidence that an expression such as 
eq. (\ref{eq:alphastable}) holds, \as representation
still provides a good modelling of many processes. 
For example, later in this work
we will show that the \as model works well even 
when the source counts do not
follow a pure power law, or when the power law is cut at a certain
flux limits.
\end{enumerate}

There is another advantage in the \as formulation:
they have been thoroughly studied in the literature,
and their properties are well understood.
Until recently the \as distributions 
were generally avoided for two main reasons: first,
the probability distribution has not a closed form
in real space (except for the particular cases of the Gaussian, Cauchy and 
Pearson distributions). This greatly hamper the development of statistical 
signal processing techniques such as maximum-likelihood and Bayesian 
estimates. The second one is that the non-Gaussian \as 
distributions have infinite 
variance (and, in some cases, as we have seen in the 
case of EPS and $\eta > 2$, a $\mu$ parameter which is not finite),
and it was considered that \as
distributions cannot be physical.
But the same objection applies to a very
well-known process, the white noise, but it is however
universally used in all fields of science and engineering. 
The variance of the theoretical white noise is not finite,
but this doesn't stop scientists to use it as an accurate
and useful model for real, finite processes. 
In the same way, \as distributions
have been shown to provide excellent 
fit to a very wide class of processes observed both
in the natural world as well as many artificial systems.
In the last few years a great deal of effort has
been carried out in the signal processing field to 
overcome the two drawbacks above mentioned.
Now there is a plethora of available methods to perform statistical
inference on \as environments. In this work we will focus on the application
of existent techniques for \as parameter extraction in order to
obtain optimal estimators of the parameters describing the differential
counts of the EPS population, namely the slope $\eta$ and the
normalisation $k$.

%__________________________________________________________________

\section{Point source parameter extraction using 
$\alpha$-stable distributions} \label{sec:parameter}

After the general discussion of \as distributions presented
in the last section, let us now return to the particular case
of the $P(D)$ and its
\as distribution posed by eqs. (\ref{eq:charf3}) to
(\ref{eq:mu}). 

According to equations (\ref{eq:alpha}) 
to (\ref{eq:mu}), the usual parameters
describing the differential counts of the EPS population 
are directly related with the parameters of the \as distribution of 
observed deflections. In particular, using
equations (\ref{eq:alpha}) and (\ref{eq:gamma}) we have that
\begin{equation} \label{eq:eta}
\eta=\alpha+1,
\end{equation}
\begin{equation} \label{eq:k}
k = \gamma \frac{ 2^{\alpha+1}\Gamma\left(\frac{\alpha+1}{2}\right)
\Gamma\left(\frac{\alpha+2}{2}\right)\sin\left(\frac{\alpha 
\pi}{2}\right)}{\pi^{3/2}\Omega_e}.
\end{equation}
Therefore, it suffices to estimate the parameters 
$\alpha$ and $\gamma$ to directly
estimate $\eta$ and $k$. Over the past years a number of efficient
estimators for the parameters of \as distributions have been developed. 
Unfortunately, most of them consider only the estimate in the case
of symmetric \as distributions ($\beta=0$), due to the fact that 
is the most common case in many signal processing applications.
On the other hand, very recently Kuruo\u glu (\cite{ercan2}) 
introduced a number of
density parameter estimators for skewed \as distributions. The simplest
estimators are based on the following idea: let us consider an
\as distribution with parameters $\alpha$, $\beta$, $\gamma$ and $\mu$,
and denote it by $S_{\alpha}(\beta,\gamma,\mu)$. 
If $X_i, \ i=1,\ldots,N$ is the sequence of data,
it is easy to show that
very simple manipulations of the data can be performed in order
to produce centred, deskewed or symmetrised sequences, respectively:
\begin{eqnarray}
X_k^C & = &  X_{3k}+X_{3k+1}-2X_{3k-2} \nonumber  \label{eq:center} \\
      & \sim & 
S_{\alpha}\left(\left[\frac{2-2^{\alpha}}{2+2^{\alpha}}\right]
\beta,\left[2+2^{\alpha}\right]\gamma,0\right) \\
X_k^D & = &  X_{3k}+X_{3k+1}-2^{1/\alpha}X_{3k-2} \nonumber \\
      & \sim &
S_{\alpha}\left(0,4\gamma,\left[2-2^{1/\alpha}\right]\mu\right) \\
X_k^S & = & X_{2k}-X_{2k-1} \nonumber \label{eq:symm} \\
 & \sim &
S_{\alpha}(0,2\gamma,0).
\end{eqnarray}
In the previous equations, the symbol $\sim$ means 
equality in  distribution
and therefore they must be regarded as
exact equations, not approximations.
The only caveat is that they work only if the samples $X_{nk}$,
$X_{nk-1}$, etc are independent random variables. Due to the 
correlations introduced by
the beam at small scales, this is not true if the samples that appear in each
individual 
summation are neighbours. Fortunately, at scales larger
than a few arcmin, where the effect
of the beam is negligible, statistical independence is satisfied.
To ensure the validity of equations (\ref{eq:center}) to 
(\ref{eq:symm}) it is enough to randomly shuffle the data before
operating. Since in this paper we do only a first order statistical modelling
--we deal with the ensemble of data rather than a time or
space series-- this has no other effect on the methods 
we present here apart from 
guaranteeing the good behaviour of the previous equations.
 
Once the distribution is conveniently centred, deskewed or symmetrised, 
the techniques for symmetric \as parameter estimate can be applied.
Kuruo\u glu (\cite{ercan2}) describes several 
groups of techniques adequate for such task:
fractional lower order moment (FLOM) methods, logarithmic
moment methods and extreme value methods. Other useful methods are
based on the study of the empirical
characteristic function of the data. Kuruo\u glu (\cite{ercan2}) 
studied the
comparison between
the different techniques and showed 
that both FLOM and logarithmic methods are
very efficient in general. In this work we are going to use 
the logarithmic 
method, as it is easier to implement.

\subsection{Logarithmic moments estimators}

Let $X$ be a set of data distributed following an 
\as $\sim S_{\alpha}(\beta,\gamma,0)$.
Let us define the \emph{logarithmic moments} of the distribution
\begin{equation} \label{eq:l1} 
L_1 = \mathbf{E} \left[ \log \left| \mathit{X} \right| \right] ,
\end{equation}
\begin{equation} \label{eq:l2}
L_2 = \mathbf{E} \left[ \left( \log \left| \mathit{X} \right| - 
\mathbf{E} \left[ \log \left| \mathit{X} \right| \right] \right)^2 
 \right] ,
\end{equation}
\noindent
where $\mathbf{E}$ is the usual estimator operator. It can be 
shown (Kuruo\u glu~\cite{ercan2})
that
\begin{eqnarray}
L_1 & = & \psi_0 \left( 1 - \frac{1}{\alpha} \right) + \frac{1}{\alpha} 
\log \left| \frac{\gamma}{\cos \theta} \right| , \\
L_2 & = & \psi_1 \left( \frac{1}{2}+\frac{1}{\alpha^2} \right)
 - \frac{\theta^2}{\alpha^2} ,
% \\
%L_3 & = & \psi_2 \left( 1- \frac{1}{\alpha^3} \right), \\
%\theta & = & \arctan \left( \beta \tan \left[ \frac{\alpha \pi}{2} 
%\right] \right), 
\label{eq:theta}
\end{eqnarray}
\noindent
where  $\psi_k$ are the values
of the polygamma function
\begin{equation}
\psi_{k-1} = \left .  \frac{d^k}{dx^k} \log \Gamma(x) \right|_{x=1},
\end{equation}
\noindent
that takes values $\psi_0 = -0.57721566\ldots$,
 $\psi_1 = \pi^2/6$, $\psi_2 = 1.2020569\ldots$, etc,  and
$\theta$ is a dummy parameter 
\begin{equation} \label{eq:def_theta}
\theta = \arctan \left( \beta   \tan \left(\frac{\alpha \pi}{2} \right)\right).
\end{equation}

This leads to the following estimators:
\subsubsection{Logarithmic estimator for $\alpha$}
Apply centro-symmetrisation as given by eq. (\ref{eq:symm})
 to the observed data to
obtain transformed data. Estimate $L_2$ and then
\begin{equation} \label{eq:est_alpha}
\alpha = \left( \frac{L_2}{\psi_1} - \frac{1}{2} \right)^{-1/2}.
\end{equation}

\subsubsection{Logarithmic estimator for $\beta$}
Once $\alpha$ has been estimated, obtain a
 distribution with $\mu=0$ (for example
centring as in eq. (\ref{eq:center})), estimate $L_2$ and then
\begin{equation} \label{eq:est_theta}
| \theta | = \left( \left[ \frac{\psi_1}{2} -
 L_2 \right] \alpha^2 + \psi_1 \right)^{1/2}.
\end{equation}
Then, estimate $| \beta |$ using eq. (\ref{eq:def_theta}).
 If centring was applied, it is
necessary to transform the resulting $\beta$
 by multiplying by $(2+2^{\alpha})/(2-2^{\alpha})$.

 According to eq. (\ref{eq:beta}) for the particular
case we consider in this work, $\beta = 1$. We provided here
the expressions for the estimation of $| \theta |$ and
$\beta$ for the general case, but in the following we will
make use of our knowledge of the true value of
$\beta$, fixing it to its theoretical value $\beta=1$
instead of trying to determine it.

\subsubsection{Logarithmic estimator for $\gamma$}
Assume again that $\mu=0$ (or make it centring the 
distribution as in the previous case), estimate $L_1$
and hence
\begin{equation} \label{eq:est_gamma}
\gamma = \cos(\theta) \exp 
\left( \left[ L_1 - \psi_0 \right] \alpha + \psi_0 \right) .
\end{equation}
Take into account that if centring was applied, 
$\gamma$ should be corrected by multiplying
by $1/(2+2^{\alpha})$.

The estimate of the location parameter $\mu$ is a tricky issue. 
Although there are some
proposed methods, they usually show problems of 
convergence and applicability. Fortunately,
the determination of $\mu$ is not necessary for estimating
the parameters $\eta$ and $k$ and therefore we will obviate this issue.

Provided with the estimators (\ref{eq:est_alpha}),
 (\ref{eq:est_theta}) and (\ref{eq:est_gamma})
we are in position of determining the parameters $\eta$ and $k$.

\subsection{Point source parameter extraction
from non-ideal power law number counts} \label{sec:testing}

In the previous sections we have showed that
an extragalactic point source population whose number
counts follow an ideal power law 
leads to an \as $P(D)$ distribution.
However, a pure power law like the one in
eq. (\ref{eq:powlaw}) is an idealisation with no physical meaning.
On one hand, eq. (\ref{eq:powlaw}) diverges
when $S$ goes to $0$ and, on the other hand, the power law extends to
infinite fluxes, which is not physically realisable. 
From the point of view of astronomy, it is not
possible to find galaxies of arbitrarily high flux and,
if we are willing to avoid Olber's paradox, a minimum flux
has to be imposed as well (at least as long as $\eta >2$).

In a real observation there must
be a minimum and a maximum flux $S_{min}$  and $S_{max}$,
respectively. This leads to a truncated power law 
\begin{equation} \label{eq:powlaw_modif}
n(S) = \left\{
\begin{array}{l l r }
0            & \textrm{if} &  S < S_{min} \\
k S^{-\eta}  & \textrm{if} &  S_{min} \leq S \leq S_{max} \\
0            & \textrm{if} &  S > S_{max} 
\end{array}
\right .
\end{equation}

From the point of view of modelling point sources in a
given sky map, if the area
of the map and the number of sources (galaxies) are finite,
the maximum flux $S_{max}$ can be safely considered  as infinite,
being the probability of finding an extraordinarily bright
source negligible.

An analogous situation can be found in
the example of white noise. The theoretical white noise
power spectrum is flat in all the range from $0$ to $\infty$, 
whereas in reality it is not possible to find such a process, but
something similar with certain cuts that depends on several factors
such as the data size, sampling, etc. Nevertheless, white noise
is a very good model for instrumental noise and many other very
well known examples. 
In a similar way, we expect that the \as model will be
a good one to describe the $P(D)$ distribution originated from
a truncated power law like in (\ref{eq:powlaw_modif}), at least
for sufficiently `well-behaved' cases.

Intuition supports the choice of \as distributions
as a fair approximation to the true $P(D)$. 
When observing a finite sample of a theoretically
infinite process, if the sample is big enough 
and the process is sufficiently well-behaved, we
expect the general model to be an adequate description
of the sample. Observing a finite number of
galaxies implies a cut like in eq. (\ref{eq:powlaw_modif}),
that is,
that we do not observe infinitely bright nor
infinitely faint galaxies. In any case, the basic shape of
the $P(D)$ distribution, an asymmetric bell-shape with
a positive tail (i.e. the impulsive behaviour), 
should be preserved. The extremes of the tails will not be
completely realised, but the basic shape should be kept
over a certain range, orders of magnitude in size if
the ratio $S_{max}/S_{min}$ is great enough.

In order to test the validity of the \as approximation as well
as the performance of the logarithmic moments estimators
introduced in the last section, 
we
%Herranz et al. (\cite{yo}) 
performed
exhaustive numerical simulations,
reproducing the observation
of typical truncated power law-distributed point sources 
detected through a Gaussian beam in a variety of cases.
%In that work it was
%shown that 
We have found that
the \as model is a very good approximation when the non-Gaussian
tail of the distribution is allowed to be well-realised. 
By `good approximation' we mean that the parameters $\eta$ and $k$
can be estimated with
errors below $5\%$ by means of the logarithmic estimators presented above. 
The goodness of the \as model depends on the following
factors:

\begin{figure} 
\centering
\includegraphics[width=\columnwidth]{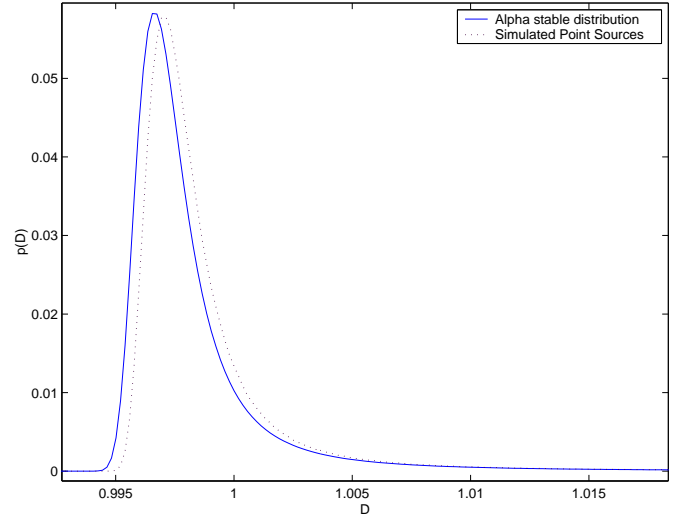}
\caption{Comparison of the $P(D)$
distribution function of simulated point sources with an 
\as distribution fitting the histogram data.
 The parameters of the EPS simulation were 
$\eta$=2.2 ($\alpha=1.2$),
$S_{min}=10^{-3}$ (in arbitrary units), $S_{max}=10^3$ 
(in the same arbitrary units), 
$N_{pix}=2048 \times 2048$, $N_{s}=N_{pix}$ 
and $FWHM=3.0$ pixels (corresponding
to a true value of the normalisation $k=3.014\times 
10^{-4}$ in the chosen arbitrary unit system). 
 The normalised histogram
of deflections is shown by means of a dotted line. 
The logarithmic
moments estimators applied to that simulation 
give the estimates $\hat{\alpha}=1.184$ and
$\hat{\gamma}=2.521 \times 10^{-4}$ (corresponding 
to an estimated value of the 
normalisation $\hat{k}=3.051\times 10^{-4}$, in the 
chosen arbitrary unit system). Using
these estimates, the corresponding \as
distribution with $\beta=1$
is shown using a solid line. The position of the dotted line has been 
slightly shifted to the right in order to make clearer the plot. }
\label{fig:sim_vs_sim}
\end{figure}

\begin{enumerate} 
\item \emph{Sample size}: there must be enough data samples to permit 
the non-Gaussian tail to appear and the logarithmic estimators to work. 
The simulations show that $\sim 10^5$ or
more samples are more than enough to work safely. Note that a typical
$512 \times 512$ pixel image satisfies this condition.
\item \emph{Flux limits}: in order to clearly observe the tail of the
distribution, the range of fluxes of the point sources must extend over
an interval big enough. In other words, there must exist a relatively few
galaxies much brighter than the vast majority of low-flux ones. A 
dynamic range 
$S_{max}/S_{min} \geq 10^4$ suffices to guarantee the good 
behaviour of the tails. This condition is satisfied in CMB 
observations as well as in many other observations at different wavelengths.
\item \emph{Slope of the differential counts}: as the distribution
becomes more and more non-Gaussian, the tails grow and there are necessary
more and more galaxies to `map' these tails. This means that for $\alpha$
values near to $2$ it is very easy to get the shape of the distribution
with few points, whereas when $\alpha$ decreases one needs a much higher
number of points and a wider flux dynamic range in order to realise the 
tails.  The results given in the two previous points are valid for 
$\alpha \in (1,1.5)$; for higher $\alpha$ values it is much easier to
realise the tails and the \as approximation holds even for smaller
number of data and tighter flux cuts. On the contrary, below $\alpha=1$
the \as approximation is less accurate unless the data size and the
flux cuts grow accordingly.
\item \emph{Beam size, pixel size and number density of sources}.
The combination of the first two factors define the so-called 
\emph{effective resolution element} of the experiment. The intuitive
idea behind the effective resolution element is that, due to the 
smoothing produced by the instrumental beam and the finite size of
the pixel, the fluctuations in the images are dumped below a certain
scale that is given by the beam and pixel sizes. If the beam is sufficiently
small, the effective resolution element corresponds to the pixel size, 
but if the beam size is bigger than the pixel size (as is usual) the
effective resolution element is related to the 
\emph{coherence scale} of the fluctuation field. For a Gaussian beam
of width $\sigma_b$ the coherence scale is roughly equal to $\sigma_b$
(Rice~\cite{rice}), and the area $\sigma_b^2$
defines the effective resolution element. 
In the case of a circular Gaussian beam $\sigma_b$ is related
the the Full Width Half Maximum (FWHM) by $FWHM=2 \sigma_b \sqrt{2\ln{2}}$.
The \as approximation is shown to work optimally
when there is approximately one source per effective resolution element.
When the number density of sources is lower, the field is 
undersampled and bigger numbers
of data are needed to realise the tails of the distribution. On the other
hand, very faint galaxies, 
whose number density is much higher than one per effective resolution
element, do not practically generate intensity fluctuations. 
In this latter case,
the fluctuations are dominated by sources whose number density is 
just below the
limit
of 1 source per effective resolution element, i.e. the brightest
undetected
ones. Hence,
the method is very sensitive to objects able to generate intensity
fluctuations, that is, the source population 
which dominates the number counts at
fluxes such that the number 
density is $\sim 1$ per effective resolution element 
or lower.
Note that this limitation is not exclusive of the \as model, but it is
shared by every other method that works with the $P(D)$ distribution
(Scheuer~\cite{scheuer2}).
The reason is that \emph{the level of $\sim$1 source per 
effective resolution element} (in the sense of coherence scale 
explained before)
\emph{roughly determines the r.m.s amplitude of the distribution}
whereas fainter sources only add some Gaussian noise 
(Franceschini et al.~\cite{franceschini1}, Barcons~\cite{barcons}).
Moreover, it is not possible to distinguish between one source per
effective resolution element
 and an infinitely large number summing up the same total
flux of an individual source.

\end{enumerate} 

Therefore, for $\alpha \in (1,2]$ the \as is a good approximation,
provided that the number of data and the flux limits of the galaxies
take reasonable values that are easily satisfied in usual
Astronomical cases.

As an example, a $2048 \times 2048$ pixel simulation was performed
in which point sources were distributed 
following a truncated power law with flux limits $S_{min}=10^{-3}$,
$S_{max}=10^3$ (in arbitrary units) and slope $\eta = \alpha+1 = 2.2$,
and then `observed' with a Gaussian beam of $FWHM=3.0$ pixels. 
The mean density of sources in the simulation was one per pixel.
The resulting distribution $P(D)$
is shown in Fig.~\ref{fig:sim_vs_sim}.
The real histogram of the deflections
produced by the EPS simulation (the solid line in
the figure) is compared with an
 \as distribution with the parameters
extracted from the EPS simulation using the logarithmic
moment estimators in eqs. (\ref{eq:est_alpha}), (\ref{eq:est_theta})
and (\ref{eq:est_gamma}). The agreement between the two curves
is very good and the relative errors in the determination of
the parameters $\alpha$ and $\gamma$ (or, conversely,
$\eta$ and $k$) are $\sim 1\%$ for both parameters (see figure caption).

Let us summarise the results of this section. We have proposed 
a set of very straightforward estimators to extract the 
parameters $\eta$ and $k$ which characterise the differential counts
of a generic point source population. These estimators,
described in equations
(\ref{eq:est_alpha}) to (\ref{eq:est_gamma}), are based on
the logarithmic moments of \as distributions and
are specifically designed to deal with non-Gaussian,
asymmetric 
{\pdf}s such as the one generated by point sources in the sky. 
From the theoretical point of view,
these estimators are efficient (Kuruo\u glu, 2001)
whereas `classical' analysis based in ordinary moments
(mean, variance, skewness and so on) is not reliable since
they do not converge.
Moreover, the estimate of the parameters $\alpha$ and $\gamma$,
directly related to the parameters $\eta$ and $k$,
is {\it direct and computationally very fast}, since it only needs the
calculation of two moments $L_1$ and $L_2$ (eqs. (\ref{eq:l1})
and (\ref{eq:l2})).
The analysis of a $2048 \times 2048$ pixels takes around one second
in a PC with a XEON 2.0 GHz processor. An approach based on `classical'
moments would require the calculation of an higher number, $N \geq 3$, of
moments, their comparison with a precalculated set of
values given by a certain model and, finally, finding the best parameters by
means of a fit (and so on).
The \as assumption works well for the case of truncated EPS distributions,
provided that the number of data is large enough and the
ratio between the cuts $S_{max}/S_{min}$ allows the tails of the distribution
to be correctly realised.
By using logarithmic estimators the parameters $\eta$ and $k$ can be
estimated with very small relative errors ($\sim 5 \%$) for a wide
range of $\eta$ values.
As it happens with other existent
methods on the $P(D)$ distribution, this method works optimally when
the average number of sources per effective resolution element is $\sim 1$.
This means that when we estimate the parameters $\eta$ and $k$ of the
differential source counts we are, actually, estimating the parameters of the
source population which dominates the counts in the flux interval around
the $S$ value corresponding to $\sim1$ source per effective resolution unit.

\section{Point source parameter extraction
from non-ideal power law number counts: 
realistic galaxy population models} \label{sec:Planck}

In section~\ref{sec:testing} 
we have established the performance and the range of applicability 
of estimators based on the \as modelling of the $P(D)$ distribution
for truncated power laws, that are a fair approximation
to the observed number counts.
Provided that reasonable conditions of applicability of the estimators
are satisfied, the results are valid for diverse fields of Astronomy,
including the study of the X-ray 
background and the modelling of unresolved point sources at
radio wavelengths. 

In this section we will go one step further and
consider state-of-the-art realistic number counts models.
As an example, we will focus on
the application of the \as
distributions to microwave observations and, in particular, 
to the images that the future ESA's Planck mission will produce.

The study of real microwave images differ from the study
of the simulations in the last section for two main reasons.
First, number counts of extragalactic sources do not follow a pure
power law distribution, but a more complex behaviour that depends
on the emission properties of galaxies, i.e. their energy spectra,
as well as their local densities
and redshift evolution. The power law distribution is only
a first order approximation to the real one. Second, microwave
images contain not only EPS signal, but also CMB radiation, other
Galactic and extragalactic foregrounds (synchrotron, free-free, dust emission
and Sunyaev-Zel'dovich effect) and instrumental noise.
All the above has to be taken into account in a realistic analysis.
In this section we will deal with the real number count 
distribution of the sources. Some hints about how to deal
with signal mixtures will be introduced in next section.

\subsection{Extragalactic point sources at Planck frequencies}
\label{sec:psplanck}

To test the efficiency of \as distributions in estimating the relevant
parameters, $\eta$ and $k$, of source number counts in CMB maps it is better
to rely on realistic cosmological evolution models for sources.
The relevant source populations at microwave frequencies are
``flat"--spectrum compact radio sources, selected at cm wavelengths,
and galaxies whose emission is dominated by dust, i.e. high
redshift spheroids and low redshift starburst and spiral galaxies,
observed in the far--IR bands. By exploiting all the available data
on extragalactic sources coming from surveys at cm and far--IR wavelengths,
T98 presented a phenomenological evolution model which allowed
to predict source number counts in the whole frequency range around the CMB
intensity peak. A thorough study on source contributions 
to the intensity fluctuations of the CMB was also presented in that paper.
As discussed in Sect.~\ref{sec:intro}, 
the observations of the microwave sky provided
by NASA's WMAP satellite 
(Bennett et al.~\cite{bennettb}) and the surveys coming
from VSA and CBI experiments, strongly support the predictions
on number counts of EPS discussed by T98, at least up to
frequencies $\nu\sim 40$ GHz. Given that ``flat''--spectrum compact sources
are the dominant population in this frequency range, we may confidently rely
on the T98 model for simulating point sources in the sky up to
$\nu\simeq 100\div 200$ GHz. On the other hand, at frequencies
$\nu\geq$ 300--400 GHz many new data have been published since 1998
and most recent evolution models fit better than the T98 model the available
data on source counts.   These recent 
models (e.g., Granato et al.~\cite{granato},~\cite{granato2})
show, in
particular, a steeper slope of the differential counts of EPS at 
$S\sim 10-100$ mJy and for $300 \leq \nu \leq 900$ GHz, where the
contribution of high-redshift spheroids show up 
(see, e.g., Perrotta~\cite{perrota}). 
On the other hand, at fluxes $S \lesssim 
0.1$ mJy all models must converge to a sub-euclidean slope 
in order to
not exceed the integrated far-IR background.
Thus, in all those flux ranges where there are no great changes of 
slope, the power-law approximation still holds and the $\alpha$-stable 
method can be effectively applied. This is the case of all (or almost)
all Planck channels given the current flux limits for source detection
foreseen for the mission (see Vielva et al.~\cite{vielva}, where
all the foregrounds have been taken into account).
On the other hand, the method could reveal not easily appliable, or not
appliable at all, to the $P(D)$ data that will come from the 
future surveys of the Herschel mission. In fact, current estimates on 
the sensitivity limits foreseen 
for Herschel (Negrello et al.~\cite{negrello})
show that, inside the flux 
range where the method could be able to recover the parameter $k$ and 
$\eta$ --i.e., for fluxes at the level of $\sim$ 1 source/beam, see 
Sect. 4.2--, EPS counts should show a sudden upturn, due to either the 
strongly positive k-correction of dust emission spectra and to the 
strong cosmological evolution of high-redshift star-forming spheroids.

In this first application of the method we still adopt the original T98 
model since we are mostly interested in testing the method rather than 
using it in a specific observation scenario.

\subsubsection{Radio sources}

Radio loud AGNs (radio galaxies, quasars and
BL-Lacs) are expected to dominate the counts in 
Planck LFI channels at fluxes $S \geq$1-10 mJy.
At frequencies around 30 GHz the 
typical values for the power law slope 
are $\eta \sim 2.0 - 2.15$ (Taylor et al.~\cite{taylor}, 
Mason et al.~\cite{mason})
at fluxes $10 \leq S(\textrm{mJy}) \leq 300$.
At higher fluxes, the data coming from classical radio surveys at cm
wavelengths show typical slope greater than the Euclidean one (i.e.,
$\eta > 2.5$. On the other hand, at lower fluxes the power law index
should keep $\sim 2.0 -  2.2$ down to fluxes of $S\sim$ a few $\mu Jy$
where it has to break down to lower values, for not exceeding current
limits on the integrated extragalactic background (see, e.g., Haarsma \&
Partridge~\cite{haarsma}, and references therein).
The number of expected detections at the 30 GHz
Planck channel, based on the T98 models
and using the Mexican Hat Wavelet detection technique (Vielva et al.~\cite{vielva}),
varies from $\sim 1800$ (when the emission of the rotational dust
is taken into account in the simulations) to $\sim 2700$ (when it is not).
Evidently, the number of detections depends on the flux detection limit
attainable by the chosen technique.

\subsubsection{Dusty galaxies}  \label{sec:dusty} 

Both `normal', i.e. spiral-like, and active galaxies show dust emission that
quickly dominates over the radio emission at wavelengths
shorter than a few mm. From a theoretical point of view, the physical
processes that govern galaxy formation and evolution are
poorly known, but there is evidence of
strong cosmological evolution in the far-IR/mm region, particularly for
early type galaxies (see Granato et al.~\cite{granato} and references therein). 
Therefore, it is not easy to model number counts of these source populations. 
SCUBA, MAMBO and IRAM surveys are rapidly providing a great amount of
data in this particular energy domain and all these data are guiding
the predictions on source counts and related statistics by means of
phenomenological as well as physical evolution models
(Toffolatti et al.~\cite{T98}; 
Guiderdoni et al.~\cite{guiderdoni};
Granato et al.~\cite{granato}, Rowan-Robinson~\cite{rowan}).
Anyway, all these models predict that number counts of EPS
are dominated by dusty galaxies at $\nu \geq 300$ GHz.
The number of these sources detectable
by Planck is variable, depending on the emission properties of
the cold dust, on the cosmological evolution of sources and
on the capability of detection techniques. Current
estimates by Vielva et al. (\cite{vielva}) based again on the T98 model with
galaxies whose positions in the sky are Poisson-distributed, 
predict the detection of $\sim 12700 $ (85 \% completeness level) point sources 
in the 857 GHz Planck channel.

\subsubsection{Total counts} \label{sec:totalcounts}

Taking into account the mixture of the different types of
galaxies and all the observational and theoretical constraints 
on them, it is evident that the number counts can not be described
by a single power law as in eq. (\ref{eq:powlaw}). Even a truncated
power law as in eq. (\ref{eq:powlaw_modif}) is not
a correct representation of the number counts. The real 
number counts need a slope $\eta$ that changes depending on the
flux range considered. 
In some cases, the data can still be approximated by the sum of
two or more populations whose number counts can be described by simple
power laws with different slopes.

In spite of this, the \as model can still be useful to
extract information about the general behaviour of the
dominating point source population from the analysis of the
total $P(D)$. In order for this affirmation to be true, the shape of the 
number counts curve should be similar to a power law at least
in the flux range of the dominant\footnote{By `dominant'
we mean the sources that produce the greatest contribution
to the number counts in the flux interval that corresponds
to number densities $\sim 1$ per effective resolution element, 
as explained in Sect.~\ref{sec:testing}.}
point sources, and the
$P(D)$ should be not much different from an \as. 
Let us
see if any of these conditions are satisfied by Planck observations.

\begin{figure} 
\includegraphics[width=\columnwidth]{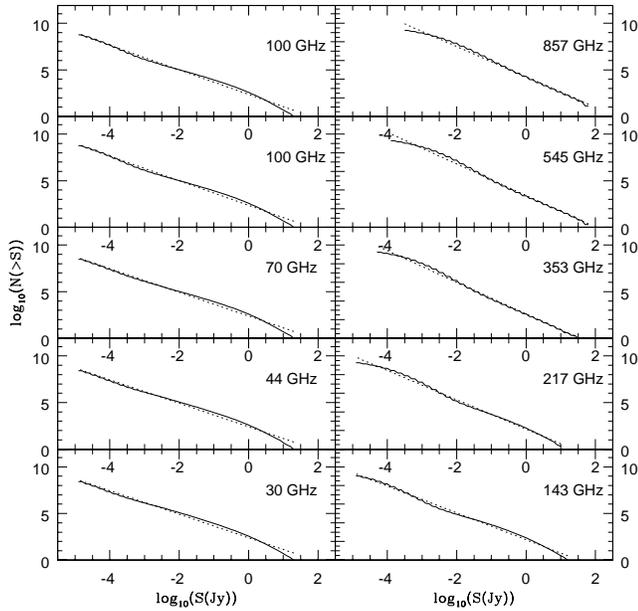}
\caption{Predicted integral counts in the ten original Planck channels. 
  The solid line shows the total number counts 
  in the simulations (including both
  radio selected and far-IR galaxies). The dotted 
  line is the best-fit single power
  law model for each channel.
  The fit is performed using only  number counts 
  of bright sources down to the flux at which
  there is approximately one source per
  effective resolution element, that is, the flux interval in which
  the statistical analysis of the $P(D)$ is sensible to the 
  number counts law.}
\label{fig:counts}
\end{figure}

Figure~\ref{fig:counts} shows the $logN$--$logS$
curves for the ten Planck channels according to the T98 model.
The solid lines show the total number counts in the
simulations (including both radio selected and far-IR galaxies).
The figure shows that the behaviour of the number counts
does not correspond to a single power law.
As discussed before, this is not surprising,
given that the spectra and the evolution properties of sources dominating
radio and far--IR source counts are quite different.
However, 
a power law fitted to the number counts (straight dotted lines), 
only at fluxes greater than the one where 
there is approximately only one source per 
effective resolution element,
does not depart much from the true curve.
The slopes of these fits lead to best-fit slopes $\eta \sim 2.1 - 2.7$. 
The best-fit slopes and normalisations calculated for each channel are 
shown in table~\ref{tb:tb1}.
According to Fig.~\ref{fig:counts},
The best-fit slopes could be considered as a good approximation to the
true slope of the source population which dominates the counts
in the relevant flux interval (see Sect.~\ref{sec:testing}) 
in each case:
i.e., ``flat''--spectrum radio sources in the low frequency channels,
dusty galaxies in the high frequency ones.

As mentioned before, in many cases the data can be approximated by the sum
of two or more populations whose number counts can be described by
simple power laws with different slopes. If one of these populations
happen to dominate over the others in the flux range corresponding to 
$\la 1$ sources per effective resolution element, the \as model 
will be a fair approximation able to determine the parameters corresponding
to such population. If there is not a dominant population, but a number
$N_p$ of roughly equally
rich
populations with different slopes, the situation
becomes more complex  and the \as model will not be correct in general.
However, as $N_p$ increases the situation tends again to $\alpha$-stability,
due to the generalised central limit theorem (Sect.~\ref{sec:alpha}).
In this latter case, the \as will allow us to determine the average
parameters of the mixed source populations.

\subsection{Validity of the \as model}

A very simple way to see if the \as model is valid for the
Planck point source simulations is to fit 
the number counts to a single power law with certain
fit parameters $\alpha_{fit}$ and $k_{fit}$, then to 
apply the logarithmic estimators in eqs. (\ref{eq:est_alpha}),
(\ref{eq:est_theta}) and (\ref{eq:est_gamma}) to 
realistic Planck EPS simulations
in order to obtain some estimated $\alpha_{logm}$ and $k_{logm}$,
and then check the agreement between the two sets of values.

\begin{table}
  \caption[]{Comparison between the point 
source parameters $\alpha$ and $k$ extracted using 
logarithmic estimators
(denoted by the \emph{`logm'} subscript)
and the best-fit power law (denoted by the \emph{`fit'} subscript) 
for the nine frequencies of the T98 Planck EPS simulations. 
The normalisation $k$ is expressed in units of Jy$^{\alpha}$ pixel$^{-2}$.
A good agreement between 
the fit and the logm values indicates that the \as approximation is valid as
a means for estimating the slope and the normalisation of the differential
number counts of the dominating point source population. Relative errors
are calculated as $r_x = 
100\times \left| x_{logm}-x_{fit} \right| / x_{fit}$, $x$ 
representing either $\alpha$ or $k$. }
  \label{tb:tb1}
  \centering
  \begin{tabular}{l l l c l l c}
    \hline  
    \hline
    \noalign{\smallskip}
    $\nu$ (GHz) & $\alpha_{fit}$ &  $\alpha_{logm}$ & $r_{\alpha}$ 
 & $k_{fit}$ & $k_{logm}$ & $r_k$ \\
    \noalign{\smallskip}
    \hline  
    \noalign{\smallskip}
    30  &  1.22  &  1.24  & 2  & 6.63$\times 10^{-4}$  & 5.81$\times 10^{-4}$ & 12 \\
    44  &  1.18  &  1.18  & 0  & 1.77$\times 10^{-4}$  & 1.73$\times 10^{-4}$ & 2  \\
    70  &  1.16  &  1.13  & 3  & 4.36$\times 10^{-5}$  & 4.72$\times 10^{-5}$ & 8 \\
    100 &  1.18  &  1.15  & 3  & 4.20$\times 10^{-5}$  & 4.30$\times 10^{-5}$ & 3 \\	  
    143 &  1.30  &  1.37  & 5  & 2.67$\times 10^{-5}$  & 1.64$\times 10^{-5}$ & 39 \\
    217 &  1.58  &  1.67  & 6  & 4.59$\times 10^{-6}$  & 2.22$\times 10^{-6}$ & 52 \\	  
    353 &  1.83  &  1.77  & 3  & 10.54$\times 10^{-5}$ & 8.81$\times 10^{-6}$ & 16 \\
    545 &  1.83  &  1.76  & 4  & 8.19$\times 10^{-5}$  & 6.61$\times 10^{-5}$ & 19 \\	  
    857 &  1.75  &  1.69  & 3  & 6.44$\times 10^{-4}$  & 5.50$\times 10^{-4}$ & 15 \\
    \noalign{\smallskip}
    \hline  \\
  \end{tabular}
\end{table}

\begin{figure} 
\includegraphics[width=\columnwidth]{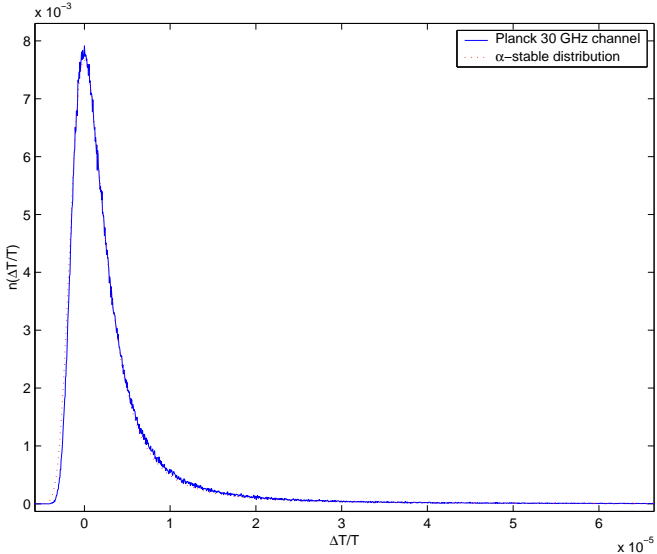}
\caption{Comparison of the histogram of the distribution
of deflections due to extragalactic point sources in the 30 GHz
Planck channel  (solid line)
and an $\alpha$-stable 
distribution (dotted line). The $\alpha$-stable
distribution has 
the $\alpha$ and $\gamma$ parameters that were estimated
by means of logarithmic estimators on the EPS data: 
$\hat{\alpha}=1.24$
and $\hat{\gamma}=7.02 \times 10^{-8}$. The EPS map is expressed
in $\Delta T/T$ units. }
\label{fig:30}
\end{figure}

We performed realistic point source simulations using the
T98 model.
The simulated sky maps were filtered using the beam sizes (circular Gaussian
approximation) and the pixel scales of the 9 Planck channels.
The number counts of the T98 model were fitted to a 
straight line in order to obtain the values of the 
parameters $\alpha_{fit}$ and $k_{fit}$ that better
fit the model in the flux range of interest.
Taking
into account the results of section~\ref{sec:testing},
the fit was performed using the number counts 
at fluxes $S>S_1$, where $S_1$ is flux over which there is approximately one
source per effective resolution element. 
The parameters
$\alpha$ and $k$ were then estimated for each channel
using logarithmic moments estimators. The results can be
seen in table~\ref{tb:tb1}. Table~\ref{tb:tb1} shows that
the validity of the \as model is better for some channels than for others.
The relative errors in the parameter $\alpha$ are
only a few percent.
The agreement in $k$ 
is worse because equations (\ref{eq:k}) and (\ref{eq:est_gamma}) 
are very sensitive to the estimated value of $\alpha$ and therefore relatively
small errors in $\alpha$ can lead to big errors in $k$. For the case of the
30 GHz channel $k_{fit}$ and $k_{logm}$ agree up to a $12\%$, and
the agreement in the 44, 70 and 100 GHz is even better,
whereas for the worst case (217 GHz) the difference between the two
values rises to a $52\%$. For the 857 GHz channel the relative difference
between $k_{fit}$ and $k_{logm}$ is $15\%$.
Therefore, we can conclude that the \as is a good model for the T98 sources,
and that the logarithmic moments estimators can recover quite well
the most representative value of $\alpha$ for the fluxes greater than 
the flux over which there is approximately one
source per effective resolution element. This value varies depending on
the channel, and corresponds in general to a few tens of mJy in the Planck case.  
The estimation of $k$ is less reliable in general, but the
method
can be still useful, however, as a means
to constrain its value.

\begin{figure} 
  \includegraphics[width=\columnwidth]{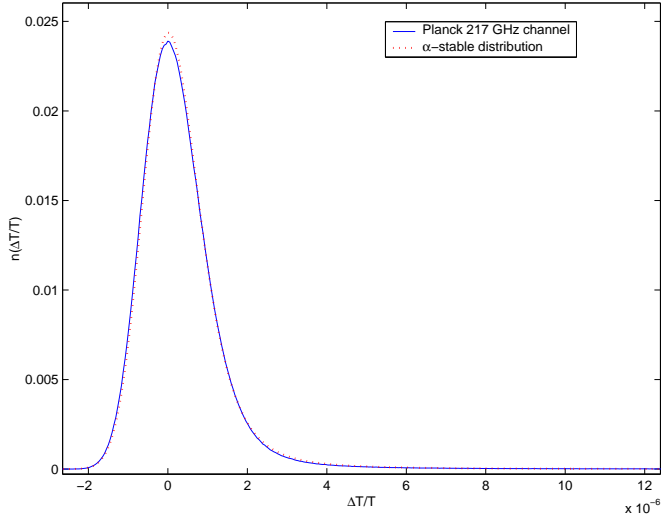}
  \caption{Comparison of the histogram of the distribution
    of deflections due to extragalactic point sources in the 217 GHz
    Planck channel  (solid line)
    and an $\alpha$-stable 
    distribution (dotted line). The $\alpha$-stable
    distribution has 
    the $\alpha$ and $\gamma$ parameters that were estimated
    by means of logarithmic estimators on the EPS data: 
    $\hat{\alpha}=1.67$
    and $\hat{\gamma}=0.34 \times 10^{-10}$. The EPS map is expressed
    in $\Delta T/T$ units. }
\label{fig:217}
\end{figure}

The results of Table~\ref{tb:tb1} 
show that the agreement between the fitted $\alpha$ and
the value estimated with the logarithmic moments
is worse for the case of the intermediate channels. This result
is easily explained taking into account
the previous discussion about the physical nature of the galaxies at microwave
frequencies. 
Whereas the low-frequency 
channels of Planck (30, 44 an 70 GHz) are 
dominated by radio-selected point sources and the 
high-frequency ones (545 and 857)
are mainly populated by far-IR galaxies, at the intermediate 
frequencies the mixing
of two main different populations distorts the shape of the 
$P(D)$ distribution,
making the \as approximation less valid.
As an example of this, 
in Figs.~\ref{fig:30}
and~\ref{fig:217}
the histogram of the deflection distribution due to the EPS as 
two of the Planck channels
are compared
with the \as distributions generated using the corresponding estimated 
$\alpha_{logm}$ and $k_{logm}$. 
Figure~\ref{fig:30} corresponds with the 30 GHz channel whereas Fig.~\ref{fig:217}
corresponds with the 217 GHz case. For this latter channel,
the basic shape of the $P(D)$ is very well approximated
by the \asb, but small differences in the region of the tail,
barely visible to the eye, are enough to hamper the validity of the \as model.
On the contrary, for the 30 GHz case in Fig.~\ref{fig:30} the agreement between the
two curves is good along all the curve, and specially good in the tail, where the most important
information is stored, leading to a good estimation of the slope $\alpha$ and the normalisation
$k$ of the radio source counts.

\section{Point source parameter extraction
  in presence of noise and other signals} \label{sec:mixing}

We have shown that the tools based on \as analysis are 
applicable to astronomical data such as those that will be observed
by the low and high frequency channels of Planck. Even when the mixture
of two roughly equally dominant different populations makes impractical
to determine the normalisation $k$ of the mixture, it is still possible
to set constraints to the global slope of the number counts. But the
previous discussion was restricted only to the emission of point
sources. 
Apart form the EPS population, the microwave sky contains emissions from
 many other astronomical 
sources (`foregrounds'), CMB radiation and instrumental noise. 
Therefore, it is not possible in general to observe the point sources
independently from the other components.
Though the addition of these other
signals does not modify the \as nature of the point sources
$P(D)$, it affects the way we can learn the parameters 
$\alpha$ and $\gamma$ from the data.
The estimate of these parameters 
can get very complicated
in presence of all these components.
In the following we will consider the simplified
(albeit very commonly found in practise)
problem of
a mixture of a Gaussian random process and EPS.
This is the case when instrumental noise (generally
modelled as Gaussian white noise) is added to the EPS signal
by the detector. In the case of CMB observations, 
that would be the case as well in `clean' sky areas (that
is, not affected by foreground contamination). We shall comment
about this  in the next section.
 
\subsection{Parameter estimation 
in Gaussian and \as mixtures}    \label{sec:withgauss}

Let us consider the case of a mixture of EPS `confusion noise' and
a Gaussian signal. 
This Gaussian signal can be instrumental noise or even 
the CMB itself if is has not been previously separated from the data.
The Gaussian
distribution is a particular case of the \as family with
$\alpha=2$. The mixture of two \as distribution with
different $\alpha$ index is not an \as distribution. Therefore,
the simple logarithmic moments estimators presented in 
Sect.~\ref{sec:parameter}
are not valid.

The characteristic function of the mixture of two independent random 
 variables can be expressed as the product of the characteristic functions
of the two original variables. That means that the characteristic function
of a mixture of an \as distribution $S_{\alpha}(\beta, \gamma, \mu)$ and
a Gaussian $\mathcal{N} ( \mathit{0} , \sigma )$ is
\begin{equation}
\psi_{mix}(w) = \exp  \left\{ i \mu w - \gamma \left| w 
\right|^{\alpha} B_{w,\alpha} + \frac{1}{2}\sigma^2 w^2  \right\} ,
\end{equation}
\noindent
where $B_{w,\alpha}$ is defined in eq. (\ref{eq:alphastable}). Applying
centro-symmetrisation as in eq. (\ref{eq:symm}) we have
\begin{equation} \label{eq:charfmix}
\psi^S_{mix}(w) = \exp \left[ -2 \gamma \left| w \right|^{\alpha}
- \sigma^2 w^2 \right] .
\end{equation}

The parameter extraction in the case of mixtures with characteristic
functions such as in eq. (\ref{eq:charfmix})
is a very difficult problem that has not been totally solved yet.
It
has been studied by Ilow \& Hatzinakos (\cite{ilow}),
who presented some  basic ideas based on the 
\emph{empirical characteristic function} (ecf)
\begin{equation} \label{eq:ecf}
\hat{\psi}_N (w) \equiv \frac{1}{N} \sum_{m=1}^{N} e^{ i w x(m)}.
\end{equation}
The ecf is a complex random variable and its expected 
value coincides with the
true characteristic function of the distribution
when the $x(m)$ samples
are i.i.d.
Ilow \& Hatzinakos (\cite{ilow}) described two 
types of methods based on the ecf to perform the estimate of the 
parameters $\alpha$, $\gamma$
and $\sigma$ in eq. (\ref{eq:charfmix}): minimum distance
 methods and moment-type methods. We tested both 
kind of methods and found that
the moment-type ones suffer from problems of stability in the particular case
under study. Therefore, we will focus on the minimum distance method.

\subsubsection{Minimum distance method} \label{sec:mindist}

In the minimum distance method, the estimate 
of the parameters $\Theta = (\alpha, \gamma,
\sigma )$ is obtained in the optimisation process
\begin{equation} \label{eq:distance}
\min_{\Theta} \int_{-\infty}^{\infty} 
\left| \hat{\psi}_N (w) - \psi_{\Theta}(w) \right|^2 W(w) dw,
\end{equation}
\noindent
where $W(w)$ is an appropriate weighting function. For example,
the choice $W(w)=\exp(-w^2)$ allows the integral (\ref{eq:distance})
to be solved by means of Gauss-Hermite quadratures, which is
computationally convenient. Thus, the estimate
of the parameters $\Theta$ is reduced to a minimisation
over three parameters. An interesting possibility appears
when one of the parameters, generally $\sigma$, is a priori known.
Then the minimisation gets much more simple.

The minimum distance method is as far as we know
the best choice present at this moment in the literature,
but it may present some problems. In particular, the
choice of the weighting function $W(w)$ is arbitrary. The
choice $W(w)=\exp(-w^2)$ is useful from the computational point
of view but it may mask important features of the ecf. We
are currently working in the elaboration of more robust 
methods, that will be the scope of future works. For the moment,
let us restrict the discussion to the minimum distance method.

\subsection{Some simple tests}

In order to test the minimum distance method
we performed a set of simulations of point sources
whose number counts follow a truncated power law,
adding a Gaussian white noise to the beam-convolved sources.
The size of the simulations was $N_{pix}=1024 \times 1024$
and the cuts of the truncated power law were set so that
$S_{max}/S_{min}=10^6$ in arbitrary units.  
The beam was set to
FWHM = 3 pixels.
For the sake of
clarity, the simulations contains
one source on average per pixel,
that corresponds  
 with $\gtrapprox 1$ source per effective resolution element
(as defined in section~\ref{sec:testing}), given the value 
of the FWHM chosen.

\begin{figure} 
\includegraphics[width=\columnwidth]{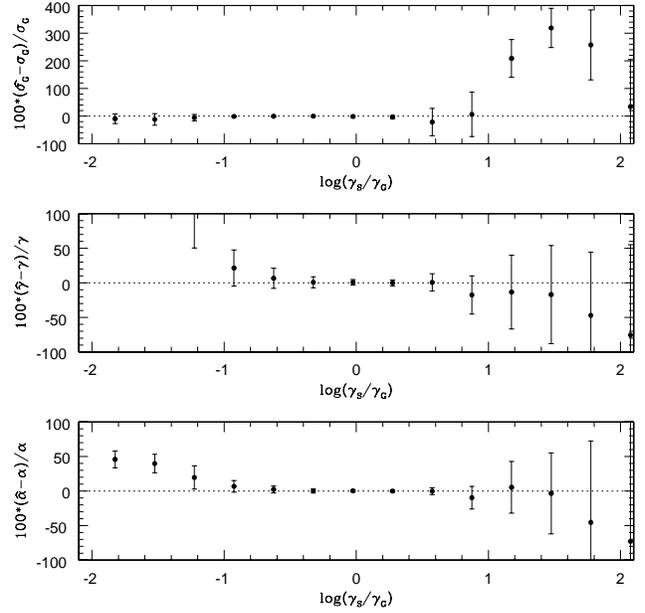}
\caption{Performance of the minimum distance method
as a function of the relative contribution of the mixture components.
The top panel shows the relative error in the determination 
of the Gaussian dispersion $\sigma_G$. The relative error 
in the determination of the \as parameter $\gamma$ is shown
in the middle panel. The bottom panel shows the relative error
in the determination of the index $\alpha$. Statistical error
bars are shown in each case.}
\label{fig:mixtures_S}
\end{figure}

We have tested the performance of the method
as a function of the relative contribution of each 
component of the mixture. In order to do that,
the slope of the EPS law was set  to $\eta=2.2$ ($\alpha=1.2$).
Fifty simulations were performed for each case.
We fixed the contribution of the Gaussian noise
to $\sigma_G=1$ (in arbitrary units) and
let vary the value of the minimum source flux $S_{min}$. This
has the effect of varying the rms of the EPS contribution, i.e.
the $\gamma_S$ of the \as distribution. 
In other words, we fix the second term in the exponential of
equation~\ref{eq:charfmix} and vary the first one.
As described in section~\ref{sec:alpha}, we
write the scale parameter of the
Gaussian contribution to the characteristic function as 
$\gamma_G = \sigma_G^2 /2$ (where $\sigma_G$ is the 
Gaussian noise dispersion). Then we can study the 
performance of the method as the ratio $\gamma_S/\gamma_G$ 
varies. 
Results are shown in Fig.~\ref{fig:mixtures_S}.

The method obtains its better performance when
the contributions of the \as distribution and the Gaussian noise
are comparable in the characteristic function. 
Inside the interval $-1 \leq \log(\gamma_S/\gamma_G)
\leq 1$ the relative errors in the determination of the three
parameters are of a few percent.
When the Gaussian contribution dominates ($\log(\gamma_S/\gamma_G) < -1$)
the estimator
`sees' only a Gaussian contribution and tries to
adjust it to the three parameters by considering that the two terms
inside the exponential in eq. (\ref{eq:charfmix}) 
are identical Gaussian exponents. Therefore, 
it wrongly gives $\alpha$ a value of 2, underestimates the 
true $\gamma_G$ by a factor $\sim 50 \%$ and greatly overestimates
$\gamma_S$.

When the point source population overdominates 
the method tends to assign the central parts of the distribution
to an almost inexistent Gaussian and to fit the remaining
tail to an \as that is more impulsive than the real one, producing 
lower values of $\alpha$ than the true ones. The $\gamma_S$ parameter
is well established in this region, but the estimated $\gamma_G$ 
gets artificially high.

In many real cases, the level of the Gaussian contribution
is a priori known. For example, the noise level of an instrument
is generally well known in most experiments. If that is the case,
the minimisation (\ref{eq:distance}) can be simplified by
fixing the value of $\sigma$ (that is, $\gamma_G$). 
Then the minimisation is performed 
in the two-dimensional parameter space $(\alpha, \gamma)$. 
In this case, the estimates of the \as parameters improve
significantly in the low $\gamma_S / \gamma_G$ regime. As an example,
in the previous test experiment for $\gamma_S=7.48\times 10^{-3}$ 
(corresponding to the point $\log(\gamma_S / \gamma_G) = -1.82$ in 
figure~\ref{fig:mixtures_S}), the estimates of the parameters
when $\sigma$ was unknown are $\hat{\alpha}=1.75 \pm 0.22$, 
$\hat{\gamma_S}=0.09 \pm 0.11$, whereas if the value of 
$\sigma$ is fixed the parameters are much better estimated,
 $\hat{\alpha}=1.2 \pm 0.1$,
$\hat{\gamma_S}=7.54 \times 10^{-3} \pm 0.91 \times 10^{-3}$.
Our tests show that if $\sigma$ is well known 
it is safe to apply the method until 
ratios $\log(\gamma_S / \gamma_G) \simeq -2.2$, with few percent relative
errors  in $\alpha$ and $\gamma$. Below that threshold, the 
performance drops quickly.

In the intermediate $\gamma_S / \gamma_G$ regime the
improvement is not so spectacular: as an example, 
for $\gamma_S=0.472$ 
(corresponding to the point $\log(\gamma_S / \gamma_G) = -0.025$ in 
figure~\ref{fig:mixtures_S}),
the results were 
 $\hat{\alpha}=1.2 \pm 0.02$,
$\hat{\gamma_S}=0.476 \pm 0.02$ when $\sigma$ was unknown and
 $\hat{\alpha}=1.2 \pm 0.01$,
$\hat{\gamma_S}=0.466 \pm 0.006$ when it was known.

In the high $\gamma_S / \gamma_G$ regime the 
knowledge of $\sigma$ gives few information
and the results are practically the same when
it is known than when it is not. 
Of course, if the EPS largely dominate 
the mixing it may be a better idea just to forget about the
minimum distance method and to apply directly the logarithmic
moment estimators of section~\ref{sec:parameter}, as the data
can be regarded as an \as contaminated with a small amount of noise.
To test this idea we just applied the logarithmic moment estimators
to our simulations. The results show that if  $\log(\gamma_S / \gamma_G) \lesssim 0.6 $
the logarithmic moments fail, as expected, but over that threshold they
work very well. As an example, for $\gamma_S=7.5$ 
(corresponding to $\log(\gamma_S / \gamma_G) = 0.57$) the
minimum distance method obtained 
 $\hat{\alpha}=1.27 \pm 0.47$,
$\hat{\gamma_S}=6.48 \pm 3.46$, whereas direct application
of the logarithmic moments gives
 $\hat{\alpha}=1.19 \pm 0.01$,
$\hat{\gamma_S}=7.24 \pm 0.12$. At higher $\gamma_S / \gamma_G$ ratios
the validity of the logarithmic moment estimators is even more
justified: for 
$\gamma_S=118.6$ 
(corresponding to $\log(\gamma_S / \gamma_G) = 2.37$) the
minimum distance method obtained 
 $\hat{\alpha}=0.55 \pm 0.80$,
$\hat{\gamma_S}=22.0 \pm 22.6$, whereas direct application
of the logarithmic moments gives
 $\hat{\alpha}=1.19 \pm 0.01$,
$\hat{\gamma_S}=109 \pm 4$.
A small negative bias ($\sim -5 \%$) seems to appear in the
determination of $\gamma_S$ for high  $\gamma_S / \gamma_G$ ratios,
probably due to the presence of the Gaussian interference.

Taking all the previously said into account, the reccomendation
is to use the minimum distance method when the
expected $\log(\gamma_S / \gamma_G) \leq 1$ and just the
logartihmic moments estimator for higher \as / Gaussian ratios. 
Fixing the value of the Gaussian contribution with a priori information
allows us to determine the \as parameters safely up to 
$\log(\gamma_S / \gamma_G) \sim -2$, whereas if $\sigma$ has to
be determined the safety region finishes around $\log(\gamma_S / \gamma_G) \sim -1$.

We also performed simulations varying the slope $\eta$ ($\alpha$)
at a given $\gamma_S/\gamma_G$ (both contributions were set
so that the contribution to the characteristic function of both
components are the same). 
The estimates for the three parameters 
$\alpha$, $\gamma$ and $\sigma$ have  low relative errors ($\sim 10 \%$)
around the values $\alpha \sim 1.0 - 1.6$. Far from this region, the
estimates become worse, specially in the low-$\alpha$ regime, where
bigger simulations or larger dynamic ranges (the ratio $S_{max}/S_{min}$)
are needed in order to `map' the \as tail correctly. On the other hand,
when $\alpha$ gets too close to 2 the two distributions are almost
Gaussian and is very hard to distinguish them.

\section{Some considerations about EPS parameter 
estimation at Planck frequencies} \label{sec:future}

As mentioned before, the microwave sky as observed by
any of the many experiments that are now being carried out or
in preparation 
is a mixture of CMB, galactic foregrounds, EPS, galaxy
clusters and instrumental noise and systematics. 
The full study of the EPS' confusion noise in such a complicated
mixture is out of the scope of this work and will be
addressed in future studies. However, some considerations 
about the expected problematics and possible results 
for the future Planck mission can be hinted in base on
the results of previous sections. 

Galactic foregrounds will be a major problem for the application
of the techniques presented in this work. Foreground signature is
nor Gaussian nor \asb, and therefore neither the logarithmic
moment estimators nor the minimum distance method for Gaussian plus \as
mixtures are expected to work well under strong Galactic
contamination circumstances. Neither will do any other of the
existent methods for the $P(D)$ study.

On the other hand, in areas of the sky where the Galactic contamination
is low there would be present basically EPS, CMB and instrumental noise
(as well as some secondary CMB anisotropies due to SZ effect). 
CMB is Gaussian or near-Gaussian distributed, and therefore 
in such sky areas the minimum distance method presented in
section~\ref{sec:mindist} can be applied, just considering
the mixture of CMB plus noise as a single Gaussian component
with joint effective variance 
$\sigma_{eff}^2=\sigma_{CMB}^2+\sigma_{n}^2$. 
%This possibility has been explored by Herranz et al. (\cite{yo}) 
%for a toy model containing only the T98 sources, CMB and noise 
%in the Planck levels, showing that 
Such approach should success
for the channels where the  EPS signal is more relevant,
that is, the low frequency channels (30 and 44 GHz) 
where radio sources are dominant
and the high frequency channels (545 and 857 GHz) 
where dusty galaxies are dominant.

Unfortunately, precisely at those channels the foreground 
contribution is expected to be very high. In particular, at 
857 GHz will be practically impossible to find a single patch
of the sky where the Galactic dust emission is subdominant with respect
to CMB. Foreground-free areas of the sky could be found outside the
Galactic plane in the intermediate Planck channels, where 
EPS are expected to be so faint that they will be under
the detection threshold of the minimum distance method.

A more interesting possibility appears if the foreground
emission can be somehow removed from the data before the
analysis. As mentioned in the Introduction, there are
several different statistical component separation methods
that are able, given some amount of a priori knowledge 
about the statistical properties of the different components,
to extract them. These methods in general are not able to
deal with unresolved point sources, and normally a residual map
containing noise, EPS and some leftovers of the components
that have not been perfectly separated
is obtained as a by-product 
of the separation.
Note that most component separation assume that the non-separable
noise is Gaussian, which is not true in presence of EPS. This
may introduce errors in the separation of the foregrounds.
An interesting possibility 
that will be studied in the future
is to introduce our knowledge on the 
\as nature of EPS confusion noise in the component separation method.
 
If a perfect separation was possible, the residual map
would contain just instrumental noise and EPS.
In that ideal case and for the T98 EPS model and
the noise levels expected for Planck, the ratios
between the point source and the Gaussian noise contributions
to the residual map would be $\log (\gamma_S/\gamma_G)
= 2.77, \ 3.09, \ 3.31, \ 2.05, \ 0.75, \ 0.30, \ 0.17$ and -0.12
for the 30, 44, 70, 100, 143, 217, 353, 545 and 
857 GHz channels, respectively.
According with the results in section~\ref{sec:mixing},
such ratios should allow the methods presented in this work
to obtain the \as parameters of the $P(D)$
with small errors for all the channels.

The real case will not be so optimistic as the one just
mentioned, but not so pessimistic as the case in which 
all the foregrounds are fully present. The study of
the performance of the \as methods 
after the application
of a component separation method
--such as the Maximum Entropy method 
(Hobson et al.~\cite{hobson}), for example-- will be carried out in
a future work.

\section{Conclusions} \label{sec:conclusions}

In this work we introduce the formalism of \as
distributions as a useful tool for the modelling and the statistical
study of number counts of undetected point
sources in astronomical
images.
When the number of faint point sources
in an image is large, the unresolved EPS contribution
creates a `confusion noise' that can be studied to obtain
information about the parent EPS population.

We have shown that when the 
differential number counts of point sources
follow a simple power law
the characteristic function of the resultant distribution
of intensity (or temperature) fluctuations
is exactly an \as one. In Sect. \ref{sec:alpha}
we briefly review the many definitions and properties
concerning the \as formalism. The \as model allows
us to model non-Gaussian, impulsive 
distributions in a flexible way and with a 
reduced number of parameters. The mathematical
foundation of \as analysis is well established
and useful results such as the generalised central
limit theorem are available. The \as distributions include
the Gaussian distribution as a particular case.

We have shown as well that, under reasonable conditions,
the \as model is well suited to describe point source
populations following truncated power law number counts.
We also have shown that even if the number distribution
in flux is not a pure power law the \as model may be
useful to estimate the parameters of the dominant source
population. In particular, 
the \as model 
has been proved to be useful and very efficient
in recovering the $k$ and $\eta$ parameters 
of the EPS number counts foreseen for the highest and the lowest 
frequency channels of the Planck surveyor mission, where point sources 
give the dominant contribution to CMB fluctuations.
As the EPS number counts in these channels are strongly dominated by
flat-spectrum radio sources (at low frequencies, LFI channels) and by
dusty galaxies (at high frequencies, HFI channels), the method discussed
here has proved helpful in determining the main parameters of the 
differential number counts  of both populations.

The \as model allows us to design efficient and computationally fast
estimators to extract the physical parameters of the EPS
populations from $P(D)$ distribution.
In particular, the logarithmic moments estimators are able to
determine the slope $\eta$ and the normalisation $k$ of the
EPS population with relative errors of a few percent over
a wide range of conditions. The method is shown to work
optimally in the case in which there is approximately
one source per effective resolution element.

Even when the EPS signal is mixed with Gaussian interference,
it is possible to estimate its parameters 
using the empirical characteristic
function, given by eq. (\ref{eq:ecf}).
We suggest a minimum distance method that is able
to deal with the presence of Gaussian noise in a wide
range of values of the ratio $\gamma_S / \gamma_G$.

Furthermore, the method uses all the information content of the data,
taking into account bright sources as well as very faint ones which 
contribute to the confusion noise distribution. 
Anyway, as discussed in section~\ref{sec:testing},
the \as model works optimally in
the case of approximately one source per effective
resolution element and, thus, its
maximum efficiency is reached at fluxes corresponding
to this source density.
An interesting possibility is to complement the
information extracted with this technique with 
other methods. For example, by detecting and counting 
bright sources it is possible to obtain the slope
$\eta$ and the normalisation $k$ in the high flux
range. These quantities can be compared 
with the ones obtained by this method
for studying the differences among source populations
which dominate the counts at high
and intermediate fluxes.
The way to proceed should be detecting first the bright
  resolved sources and, once removed these sources from the data --for example
  masking the area occupied by them--, analysing the $P(D)$ of the
  unresolved sources by means of the statistical estimators introduced in this
  work. The statistical study of the $P(D)$ allows us to go
  below the detection threshold. The flux limit these methods can reach
  depend strongly on the characteristic of the experiment: in absence
  of noise and other foregrounds, the theoretical limit is the flux
  corresponding to approximately one source per effective resolution element.
  In presence of noise and foregrounds, this limit will be higher and
  must be determined for each particular case.

As mentioned before,
the \as model works well even 
there is the presence of two or more source populations
with different slopes of the number counts, provided
that the departure from a simple power-law is not extreme
in the relevant flux range.
Then, techniques such as the logarithmic moments
estimators are able to estimate a single slope
that corresponds to the one 
of the dominant source population.
It is possible, however, to refine
the results. A method similar to the minimum
distance method presented in Sect.~\ref{sec:mindist}
can be conveniently modified to include 
more than one different \as components in order to 
determine the parameters $k$ and $\eta$ of two or, possibly,
more source populations.

Work to obtain an optimal technique to deal with
\as mixtures is currently in progress.

In this work we considered that the 
spatial distribution of the sources in the sky
is uniform. However, the sources are expected to show
some degree of autocorrelation. Source clustering 
will produce a broadening in the $P(D)$ distribution
(Barcons~\cite{barcons}). This effect should be taken into account
in a future work. We did not take into account
the presence of Sunyaev-Zel'dovich (SZ) effect in the
CMB maps. The signature of SZ clusters is expected to be
similar in nature to the signature of EPS\footnote{as far as the 
two populations are observed through
large effective aperture beam so that clusters and EPS can be 
considered as point-like objects}
and, thus, the
formalism of this work could be applied to it. At
frequencies $\nu < 217$ GHz the SZ effect will have a
negative contribution to the maps, producing
skewed $P(D)$ distributions with negative tails. 
Several ways of discriminating between unresolved point sources and
SZ clusters in CMB maps have been studied by 
Rubi\~no-Mart\'\i n \& Sunyaev (\cite{rubi}).

The previous conclusions can be applied to different fields
in Astronomy including the X-ray background, radio Astronomy 
and, in general, to all the observations in which it is interesting
the study of the statistical properties of undetected point sources.

The potentialities of the \as modelling go further than
the designing of estimators such as the ones presented
in  this work. The \as model allows us to
use  techniques present in the signal processing literature
for achieving a complete probabilistic
description of the data that can be used for 
more ambitious goals such as Bayesian estimation, denoising,
etc. Future works will explore these interesting possibilities.

%__________________________________________________________________

\begin{acknowledgements}
We thank the anonymous referee for carefully reading the
manuscript and for helpful suggestions.
DH and LT acknowledge partial financial support from the
European Community's Human Potential Programme
under contract HPRN-CT-2000-00124 CMBNET. 
We thank  E. Salerno, A. Tonazzini,
E. Mart\'\i nez - Gonz\'alez and P. Vielva for useful 
comments.
This work has used the software package HEALPix
(Hierarchical, Equal Area and iso-latitude pixelisation of the
sphere, http://www.eso.org/science/healpix), developed by
K.M. Gorski, E. F. Hivon, B. D. Wandelt, J. Banday, F. K.
Hansen and M. Bartelmann.
\end{acknowledgements}

\end{document}